\documentclass[lettersize,journal]{IEEEtran}
\usepackage{amsmath,amsfonts}
\usepackage{algorithmic}
\usepackage{algorithm}
\usepackage{array}
\usepackage[caption=false,font=normalsize,labelfont=sf,textfont=sf]{subfig}
\usepackage{textcomp}
\usepackage{url}
\usepackage{verbatim}
\usepackage{graphicx}
\usepackage{cite}
\usepackage{booktabs}
\usepackage{tablefootnote}
\usepackage{threeparttable}
\usepackage{xcolor}
\usepackage{soul}
\newcommand{\comm}[1]{}
\usepackage{bm}
\newcommand{\red}[1]{\textcolor{red}{#1}}
\usepackage{calrsfs}
\begin{document}

\title{FatigueFusion: Latent Space Fusion for Fatigue-Driven Motion Synthesis}
\author{Iliana Loi and
Konstantinos Moustakas~\IEEEmembership{Senior Member,~IEEE}
\thanks{Iliana Loi and Konstantinos Moustakas are with the Department of Electrical and Computer Engineering, University of Patras, Patras, 26504 Greece}
\thanks{Corresponding Author: loi@ceid.upatras.gr}
\thanks{\textbf{This work has been submitted to the IEEE for possible publication. Copyright may be transferred without notice, after which this version may no longer be accessible.}}
}

\markboth{Journal of \LaTeX\ Class Files,~Vol.~.., No.~.., ..~2026}%
{Shell \MakeLowercase{\textit{Loi et al.}}: FatigueFusion: Fatigue Feature Fusion in Latent Space for Fatigue-Driven Motion Synthesis}


\maketitle

\begin{abstract}
Investigating the impact of fatigue on human physiological function and motor behavior is crucial for developing biomechanics and medical applications aimed at mitigating fatigue, reducing injury risk, and creating sophisticated ergonomic designs, as well as for producing physically-plausible 3D animation sequences. While the former has a prominent position in state-of-the-art literature, fatigue-driven motion generation is still an underexplored area. In this study, we present \textit{FatigueFusion}, a deep-learning architecture for the fusion of fatigue features within a latent representation space, enabling the creation of a variation of novel fatigued movements, intermediate fatigued states, and progressively fatigued motions. Unlike existing approaches that focus on imitating the effects of fatigue accumulation in motion patterns, our framework incorporates algorithmic and data-driven modules to impose subject‑specific temporal and spatial fatigue features on non-fatigued motions, while leveraging PINN-based techniques to simulate fatigue intensity. 
Since all motion modulation tasks are taking place in latent space, FatigueFusion offers an end-to-end architecture that operates directly on non-fatigued joint angle sequences and control parameters, allowing seamless integration into any motion synthesis pipeline, without relying on fatigue input data. Overall, our framework can be employed for various fatigue-driven synthesis tasks, such as fatigue profile transfer and fusion, while it also provides a solution for accurate rendering of the human fatigue state in both animation and simulation pipelines.        
\end{abstract}


\section{Introduction}
\IEEEPARstart{H}{uman} fatigue is widely explored in recent literature, with works focusing on both the alterations in muscle physiology, and internal body and cognitive processes under fatigued conditions \cite{Lock2018, Mahdavi2024, Pernigoni2024}, as well as the effects of externally perceived fatigue in human movement patterns and posture \cite{Zhang2022, Cortes2014}. Works in this domain aim to develop fatigue mitigation and prevention strategies in work environments \cite{Dhawan2025, Gander2017}, sports activities \cite{Thorpe2017} as well as digital interfaces \cite{Wang2025investigating}, and ergonomic risk assessment designs \cite{Lorenzini2019}. In parallel with traditional biomechanics studies, data-driven methods evolving around fatigue are primarily utilized to solve classification tasks, aiming to identify either physical \cite{Hwang2025, Aoki2021}, or mental fatigue \cite{Pavel2023} from human locomotion, without accounting for the modeling of fatigue. 

Regarding animation frameworks, state-of-the-art approaches range from traditional Machine and Deep Learning (ML/DL) models, such as Recurrent Neural Networks \cite{Carneros2024} for motion prediction and synthesis, to more recent Transformer-based \cite{Hou2023, Wan2024} and Diffusion-driven approaches \cite{Zhang2024, Gao2024} capable of generating text‑conditioned motion sequences. Such methods are being optimized to address performance, computational efficiency, and predictive accuracy issues, while achieving motion variability and realism. Furthermore, physics-based motion synthesis techniques incorporate biomechanical \cite{Kang2025} and physical \cite{Li2025} domain knowledge to constrain and penalize generated motions, thereby ensuring closer adherence to physically realistic motion sequences.

Even though fatigue is essential for both developing biomechanical engineering applications and generating realistic physics-based animation, modeling fatigued movements remains an area where very little to no research is conducted, with fatigue-driven motion synthesis techniques being limited to two works \cite{Loi2025, Cheema2023}. Another limitation is that the latter works implement deep learning architectures designed to replicate the impact of fatigue accumulation in generated motions, neglecting to reflect diversified fatigued characteristics. 

Addressing these research gaps, we present \textit{FatigueFusion}, an end-to-end data-driven framework for fusion of fatigue-related features in latent space, to produce a variety of novel fatigue motions, intermediate fatigued states, and progressively fatigued movements. FatigueFusion comprises of three complementary modules, each one functioning on one of three dimensions: namely time, space, and fatigue intensity. The \textit{Fatigue Tempo} module uses interpolation-like techniques to extract temporal fatigued features per stance phase (e.g., gait de-synchronization), whereas the \textit{Fatigue Features} model encodes and fuses spatial fatigued features, i.e., the effects of externally perceived fatigue in movement (e.g., leg instability, extreme lumbar bending, etc.) into a latent representation by leveraging a synergy between a Conditional Variational Autoencoder (CVAE) and a simple Autoencoder (AE). These subject‑specific fatigue characteristics, derived from the DUO‑Gait fatigued dataset \cite{DUO_GAIT} used to train the framework, are hereafter referred to as each subject’s \textit{fatigue profile}. To model the progressive accumulation of fatigue, we employ an extended version of 3CC-$\lambda$ (i.e. \textit{Fatigue Intensity} module), introduced in our prior work \cite{Loi2025}, which represents a PINN adaptation of the Three-Compartment-Controller (3CC) Model \cite{Liu2002, Xia2008, Frey-Law2012}. Overall, during inference, the mapped temporal features, latent spatial representations, and the fatigue‑intensity scaling are fused to synthesize a unique fatigued motion sequence. 
The contributions of this work are summarized as follows:
\begin{enumerate}
    \item Introduction of FatigueFusion, a deep learning framework for fatigue-driven motion generation. By sampling fatigue features directly from latent space, our model enables the synthesis of fatigued motion sequences exhibiting various spatial and temporal fatigue characteristics, including intermediate fatigue states and motions that progressively transition from non‑fatigued to fatigued conditions.
    \item FatigueFusion is independent of the animation framework supplying its input, and thereby can be integrated into any motion synthesis framework. Furthermore, by embedding fatigue representations in latent space, our model bypasses the need for fatigued motion data during inference, relying solely on non-fatigued joint angle sequences and control variables, which enable for the adaptation of subject-specific motion parameters.
    \item Our approach complements state-of-the-art works that focus on developing methods for simulating fatigue accumulation by allowing the modulation and fusion of temporal and spatial features that describe the impact of fatigue on human movement patterns.
\end{enumerate}

\section{Related Work}

\subsection{Data-Driven Animation}
\label{sec:animation}

Researchers and developers in the domains of graphics, animation, and gaming develop both deterministic and probabilistic DL methods that exploit human motion‑capture data and movement histories, such as prior pose sequences or joint angle trajectories, to generate or predict the poses and joint motions of 3D humanoid characters \cite{Loi2023_review}. Deterministic data-driven approaches have employed a wide spectrum of neural network architectures, spanning from conventional Feed-Forward \cite{Holden2016} models to advanced recurrent models \cite{Zhou2021}, with the aim of synthesizing motion sequences that converge toward a single deterministic trajectory, typically regressing to the mean pose of the ground‑truth motion (i.e., the motion sequences used during training). 

On the other hand, probabilistic motion synthesis methods dominate the current literature, since they inject stochasticity in the generation process by fitting a Gaussian distribution to the latent distribution of the motion, diversifying the resulting movement, and ultimately producing all plausible pose sequences of a virtual character based on previous frames and/or control inputs. Widely used models in this area constitute Variational Autoencoders (VAEs) \cite{Petrovich2021}, Conditional Variational Autoencoders (CVAEs) \cite{Cai2021}, Generative Adversarial Networks (GANs) \cite{Li2022}, and Transformers \cite{Wang2025}. Nevertheless, the most prominent methods in this category are Diffusion models \cite{Zhang2026, Wang2025difusion, Amballa2025, Li2025_2, Zhang2024, Gao2024, Zhang2024tedi, Dabral2023}, taking advantage of the diffusion process which, during inference, involves the gradual disentanglement of a normally distributed noise vector to generate a meaningful pose sequence. Such methods, take advantage of the above DL architectures to incorporate stochasticity and condition their results on multiple data inputs, such as natural language prompts, audio clips, and videos. Characteristic works in this domain are developing transformers for body-part aware motion generation \cite{Zhang2024, Dabral2023} as well as for distance-aware multi-person interactions \cite{Zhang2026}, GANs in latent space to accelerate training and inference times \cite{Amballa2025}, and extensions of Diffusion Denoising Probabilistic Models (DDPM) for enabling temporally varying denoising \cite{Zhang2024tedi}. 

\subsection{Physics-based Animation using Deep Learning}
\label{sec:physics_based_animation}
Physics-based animation refers to motion generation techniques that ground their results on physics laws and/or physical parameters (e.g., joint torques, forces, velocities, etc.) to penalize the solution space and, hence, enhance the realism of the produced motion sequences. Reinforcement Learning (RL) \cite{Xu2025, Truong2024, Yao2024} and Adversarial methods \cite{Zhang2025} are usually favored for physics-based motion synthesis tasks, since physical constraints can be effectively applied through reward/penalty functions, rendering these models capable of learning robust control policies resulting in the imitation of a wide range of physically-plausible motions.  

However, most recent works in this area combine diffusion approaches with physics optimization systems. In \cite{Li2025}, a motion generator works in collaboration with a physics-simulator that performs motion imitation constraint on physical factors, to learn to encode synthetic noisy motion data into physically-realistic pose sequences, bypassing the need for highly refined motion capture data for training. An adversarial-based module is also incorporated in this model to accelerate and stabilize the physics-guided generative process so as to produce motions devoid of artifacts. Another innovative work combining diffusion-based and physics-informed models, presented in \cite{Kang2025}, is an end-to-end architecture called BioVAE. BioVAE consists of a diffusion model and a VAE, whose encoder maps pose sequences to EMG signals, while incorporating musculoskeletal dynamics priors in the latent space to infer joint-specific rotational accelerations. Based on these accelerations, the VAE's decoder produces residuals to guide the diffusion process. Apart from creating physically-consistent motions, this biomechanics-aware framework supports and generalizes across various motion modulation tasks, such as motion generation, editing, and refinement \cite{Kang2025}. Moreover, a framework consisting of a transformer autoencoder, which encodes biomechanical parameters (i.e., joint positions, contact forces, muscle activations, etc.) into latent space, a spatial controllability model for adjustment over biomechanical motion variables, and a text-conditioned diffusion model operating upon latent space, rendering the overall architecture computationally inexpensive during both training and inference, is outlined in \cite{Tashakori2025}.

\subsection{Fatigue-Driven Motion Synthesis}

Combining motion generation data-driven models, as the ones outlined in Section \ref{sec:animation}, with human biomechanics estimation approaches relying on standard machine/deep learning models or PINNs, enables the derivation of physically-plausible and biomechanically consistent results, similar to the ones inferred from physics-based animation works (Section \ref{sec:physics_based_animation}). Although the ultimate goal of all physics-informed motion synthesis frameworks is the naturalness of motion, they do not consider the impact of fatigue while generating temporally-evolving motion sequences, allowing for the simulation of a never-ending active human state and the accumulation of non-negligible errors (e.g., not simulating the gradual reduction of maximum exerted joint torques and muscle forces after a repetitive fatiguing task). 

Addressing this limitation, there are only two works developing fatigue-driven motion synthesis exploiting neural network architectures \cite{Loi2025, Cheema2023}. Both works leverage the Three-Compartment Controller (3CC) state machine to account for fatigue-induced fluctuations modulating joint torques, and as a result model the impact of externally perceived fatigue in human motion. In \cite{Loi2025}, an end-to-end deep learning architecture incorporating semi-dynamics surrogate ID and FD models and a PINN adaptation of 3CC, deterministically models fatigue in upper limb complex movements, without taking into consideration fatigue motion data during training or inference. Fatigue-driven animation in \cite{Cheema2023} is realised by combining 3CC with a Reinforcement Learning (RL) approach for controlled motion imitation. The latter method, learns compensation techniques for overcoming the fatigue effects on motion, such as small pauses between consecutive fly-kicks, or transitioning from running to walking. Nevertheless, none of these works encode personalized fatigue characteristics in residual motions, yet they mainly focus in simulating fatigue accumulation over time. The proposed FatigueFusion framework addresses all aspects of fatigue across time, space and intensity, allowing for the adaptation of different fatigue features through control variables.

\section{Methods}

\subsection{FatigueFusion Framework}

We propose an end-to-end fatigue-specific motion generation framework consisting of $3$ modules, addressing all aspects that will render a fatigue motion realistic: i) the \textit{Fatigue Tempo} module, which leverages interpolation techniques to encode fatigued temporal features per stance phase such as an irregular (de-synchronized) walking pattern due to fatigue, ii) the \textit{Fatigue Features} module, which accounts for the fusion of subject-specific fatigued features (fatigue profiles) into latent space, to produce novel fatigued motions and intermediate fatigue states, while iii) the \textit{Fatigue Intensity} module is an extension of 3CC-$\lambda$ presented in our prior work \cite{Loi2025}, and is employed to inject fatigue to the residual motions.

\begin{figure*}[h!]
\centering
\includegraphics[width=\textwidth]{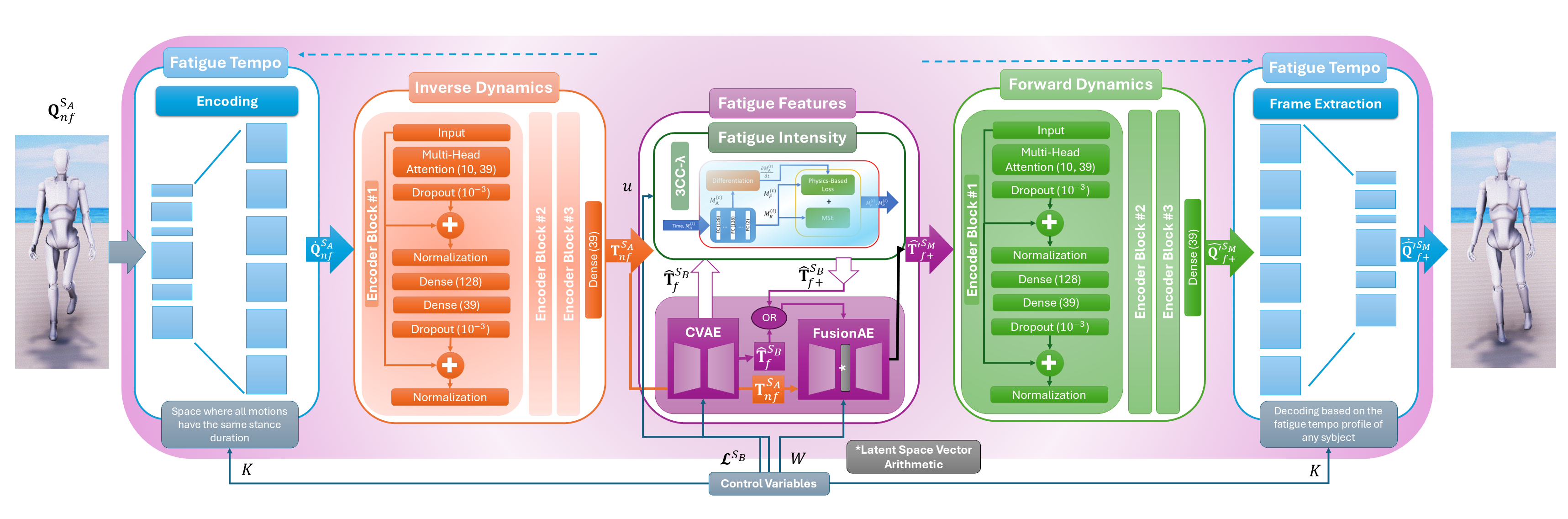}
\caption{A general overview of our FatigueFusion pipeline consisting of Fatigue Tempo, Fatigue Features and Fatigue Intensity modules. The Figure for 3CC-$\lambda$ model was reproduced from \cite{Loi2025}.}  
\label{fig:fatigue_fusion}
\end{figure*}


As illustrated in Fig. \ref{fig:fatigue_fusion}, at each time frame, our framework receives a non-fatigued motion sequence (joint angles) $\mathbf{Q}_{nf}^{S_A}$, which is first upsampled into a normalized temporal domain via the Fatigue Tempo module, acquiring uniform stance phase durations, hence producing $\dot{\mathbf{Q}}_{nf}^{S_A}$. Then it passes through a surrogate Inverse Dynamics data-driven model that accounts for the transformation of joint angles to joint torques, yielding $\mathbf{T}_{nf}^{S_A}$. The latter joint torques are given as input to the Fatigue Features module, along with control variables $\{\mathcal{L}^{S_B}, W\}$ that specify which spatial fatigue features and how they will be fused into the non-fatigued sequence. The Fatigue Intensity module is incorporated into the Fatigue Features module to amplify the impact of fatigue, thus highlighting the features of the already-fatigued torques $\hat{\mathbf{T}}_{f}^{S_B}$ (output of CVAE), and producing $\hat{\mathbf{T}}_{f+}^{S_B}$. The output of the Fatigue Features module, i.e., fatigued joint torques $\hat{\mathbf{T'}}_{f}^{S_M}$ generated via FusionAE, is fed to a Forward Dynamics deep learning model, sharing the same architecture as its ID counterpart, which performs the reverse transmutation, from joint torques to joint angles $\hat{\mathbf{Q'}}_{f}^{S_M}$. The fatigued joint angle sequence is subsequently downsampled to obtain the original pose sequence, or apply the tempo of any subject to the resulting motion. Overall, this framework enables the fusion of various temporal and spatial fatigue features, along with different levels of fatigue (e.g., entangling the fatigue tempo of a subject $S_A$ with the fatigue features of another, $S_B$), to generate novel fatigued movements.     

The ID and FD models preceding and succeeding the Fatigue Features module, respectively, are Transformer-based models trained upon the DUO-Gait dataset as well. As mentioned above, both models employ the same architecture, as depicted in Fig. \ref{fig:fatigue_fusion}, comprising of $3$ stacked encoder blocks, each equipped with a multi-head attention layer with $10$ heads, followed by fully connected layers. In contrast with our prior work \cite{Loi2025}, in which the ID and FD models were designed to be joint-specific to allow for joint-specific fatigue configurations and mitigation of joint artifacts on a joint level, the current architectures operate over \textit{all} joints simultaneously, i.e., a single model for each biomechanics task is trained with all kinematics and dynamics data. Consequently, inter-joint correlations can be extracted to improve predictive accuracy and convergence while achieving faster training times. 

\subsubsection{Motion Representation}

Motion sequences are represented as a set of joint angles $\textbf{Q} \in \rm I\!R^{TxJ+3}$ (a 2D matrix of Euler Angles) incorporating also $x, y, z$ root translations $\textbf{R} \in \rm I\!R^{Tx3}$, spanning across $T$ time frames, where $J$ is the number of joint angle features. Hereafter, we will refer to matrix $\textbf{Q}$, as "joint angle sequences" for simplicity, while in reality this matrix also contains the root translations. In addition to joint angle sequences and root translations, our framework takes as input a set of control variables $C = \{\mathcal{L}^{S}, W, K, u\}$, where $\mathcal{L}^{S}$ is a subject label and $W$ a vector of fusion weights, to allow for the adaptation of spatial fatigue features, $K$ is a vector of parameters facilitating the modulation of temporal fatigue features (e.g. interpolation stance length), and $u \in [0, 1]$ is a lower limit factor to regulate the intensity of fatigue accumulation. 

Furthermore, since both Fatigue Features and Fatigue Intensity (i.e., 3CC-$\lambda$) modules operate in torque space (see the following Sections \ref{sec:fatigue_features} and \ref{sec:fatigue_intensity} for more details), we utilized OpenSim's ID tool to infer joint torques $\textbf{T} \in I\!R^{TxJ}$ from the respective joint angles. These torques are required for the training of the surrogate ID and FD models included in our framework. The ID tool integrated into OpenSim, employs Newton's classical equation of motion Eq. (\ref{eq:L_ID}), to describe generalized forces (joint torques) as a function of generalized positions (joint angles) and their derivatives (joint velocities and accelerations): 

\begin{equation}
    \mathbf{\tau} = M(\textbf{q})\ddot{\textbf{q}} + C(\textbf{q}, \dot{\textbf{q}}) + G(\textbf{q}) 
    \label{eq:L_ID}
\end{equation}

where $\mathbf{q},\mathbf{\dot{q}},\mathbf{\ddot{q}} \in I\!R^J$ represent the corresponding vectors of generalized positions, velocities, and accelerations, $\mathbf{\tau} \in I\!R^J$ is the vector of generalized forces, $M(\mathbf{q}) \in I\!R^{NxN}$ is the mass matrix, $C \in I\!R^J$ is the vector of Coriolis and centrifugal forces, $G \in R^J$ is the vector of gravitational forces. This equation is computed for each time frame $t = \{0, ..., T\}$ of the motion, hence, we denote that $\textbf{Q} = \{\textbf{q}_{0}, ..., \textbf{q}_{T}\}^T$ and $\textbf{T} = \{\mathbf{\tau}_{0}, ..., \mathbf{\tau}_{T}\}^T$. The musculoskeletal model\footnote{Available online at: \url{https://simtk.org/projects/full_body/}}\cite{Rajagopal2016} used in both the above ID procedure and in the IMU IK process to derive joint angle sequences from raw IMU orientation data, as outlined in Section \ref{sec:dataset}, \comm{is a Hill-type full-body model presented in \cite{Rajagopal2016}}is a full-body model with a total of 37 Degrees of Freedom (DoFs). \comm{This model has a total of 37 Degrees of Freedom (DoFs) and features 80 muscle-tendon units in the lower extremities and 17 ideal torque actuators in the torso/upper limbs.}  

Ultimately, our framework incorporates both non-fatigued and fatigued generalized coordinates and force sequences during training; thus, we denote non-fatigued joint angle and torque matrices as $\textbf{Q}_{nf}^S$ and $\textbf{T}_{nf}^S$, respectively, while their fatigued counterparts are represented as $\textbf{Q}_{f}^S$ and $\textbf{T}_{f}^S$. $S$ serves as a notation for a specific subject, which will aid us to distinguish and discuss the different fatigue profiles in Section \ref{sec:results}. 

\subsubsection{Fatigue Tempo Module}


The main idea behind the Fatigue Tempo Module is the fact that human gait presents subject-specific variability in stance phase duration, rendering it an inherently personalized feature, especially in the case of fatigue, when gait becomes even more irregular. To extract these temporal characteristics, we first encode all motion sequences (joint angles) into a normalized temporal space in which all motions have the same stance duration. This is achieved through a novel interpolation algorithm, whose key feature is that it preserves \textit{all} original frames of a motion while inserting additional frames between them, ensuring that \textit{each} stance is expanded based on the duration of the longest original stance out of all subjects in DUO-Gait dataset, resulting in $300$ frames per stance in the interpolated space. This dynamic mapping of key frames of movement is realised through the "Encoding" component of Fatigue Tempo Module, as illustrated in Fig. \ref{fig:fatigue_fusion}. For example, an individual's fatigue gait with stance durations of lengths $\{76, 30, 61, 25, 71, 30\}$, exhibiting a stumbling gait (fatigue tempo profile), will acquire a gait with stance phase durations of $\{300, 300, 300, 300, 300, 300\}$ (stride regularization). 
The mapping from the original to the interpolated domain, is encoded in the "Frame Extraction" component, to facilitate the extraction of key temporal frames representing a unique temporal fatigue profile, from the normalized space. In this way, the original frame sequence or a frame sequence bearing any fatigue tempo profile, can be produced, rendering the latter component essential for our FatigueFusion framework, since it enables subject‑specific temporal fatigue characteristics to be transferred onto any motion sequence. More specifically, an encoded motion sequence can be decoded to any irregular phase tempo, allowing for various combinations of temporal fatigue features such as attributing a non-fatigued gait, with a specific temporal fatigue feature such as hobbling, rendering a fatigued gait that does not exhibit stumbling, to stumble, etc. Moreover, this module allows for performing any fatigue feature modulation task within a stance phase, hence entangling fatigue features (e.g. hunchbaking with stumbling, etc.).


\subsubsection{Fatigue Features Module}
\label{sec:fatigue_features}
\paragraph{Conditional Variational Autoencoder}

At the core of the Fatigue Features module is a Conditional Variational Autoencoder, which, during inference, is fed with a non-fatigued joint torque sequence, $\mathbf{T}_{nf}^S$ (in the mathematical analysis is referred to as $x$ for simplicity), and a condition, i.e., a label $\mathcal{L}^S$, as a subject embedding, to produce a unique fatigued movement. In simple words, CVAEs incorporate an encoder to compress input data in latent space $\mathcal{Z}$, and a decoder that reconstructs fatigued joint torque sequences based on a latent vector $z$ sampled from latent space, and a condition. Mathematically, CVAE defines Eq. \ref{eq:CVAE}:

\begin{equation}
    p_{\theta}(x |\mathcal{L}) = \int p_{\theta}(x |z, \mathcal{L})p(z|\mathcal{L})dz
\label{eq:CVAE}
\end{equation}

with latent variables of training data being expressed as a probability distribution $p(z|\mathcal{L})$ (usually = $N(0, I)$, prior), and $\theta$ represents the decoder parameters. Since 
the integral is \comm{rendered}intractable, the encoder approximates the posterior distribution $q_{\phi}(z|x,\mathcal{L}) = N(\mu_{\phi}(x, \mathcal{L}), \sigma_{\phi}(x, \mathcal{L}))$, to facilitate the learning of the condition distribution $p(x|z, \mathcal{L})$ by the decoder. This probabilistic encoding is achieved by injecting Gaussian noise $\epsilon \sim N(0, I)$ in the latent vector sampling process as mandated by Eq. \ref{eq:z} (re-parameterization trick):
\begin{equation}
    z = \mu_{\phi}(x, \mathcal{L}) + \sigma_{\phi} (x, \mathcal{L}) \epsilon 
\label{eq:z}
\end{equation}

where $\mu_{\phi}(x, \mathcal{L})$ and $\sigma_{\phi}(x, \mathcal{L})$ are, respectively, the encoder's estimate of the posterior mean and standard deviation of the latent variable $z$, given the input $x$ and condition $\mathcal{L}^S$. In our case, each latent vector $z$, represents a latent representation of personalized fatigued features, based on which a novel movement will be produced.   

\begin{figure*}[h!]
\centering
\includegraphics[width=\textwidth]{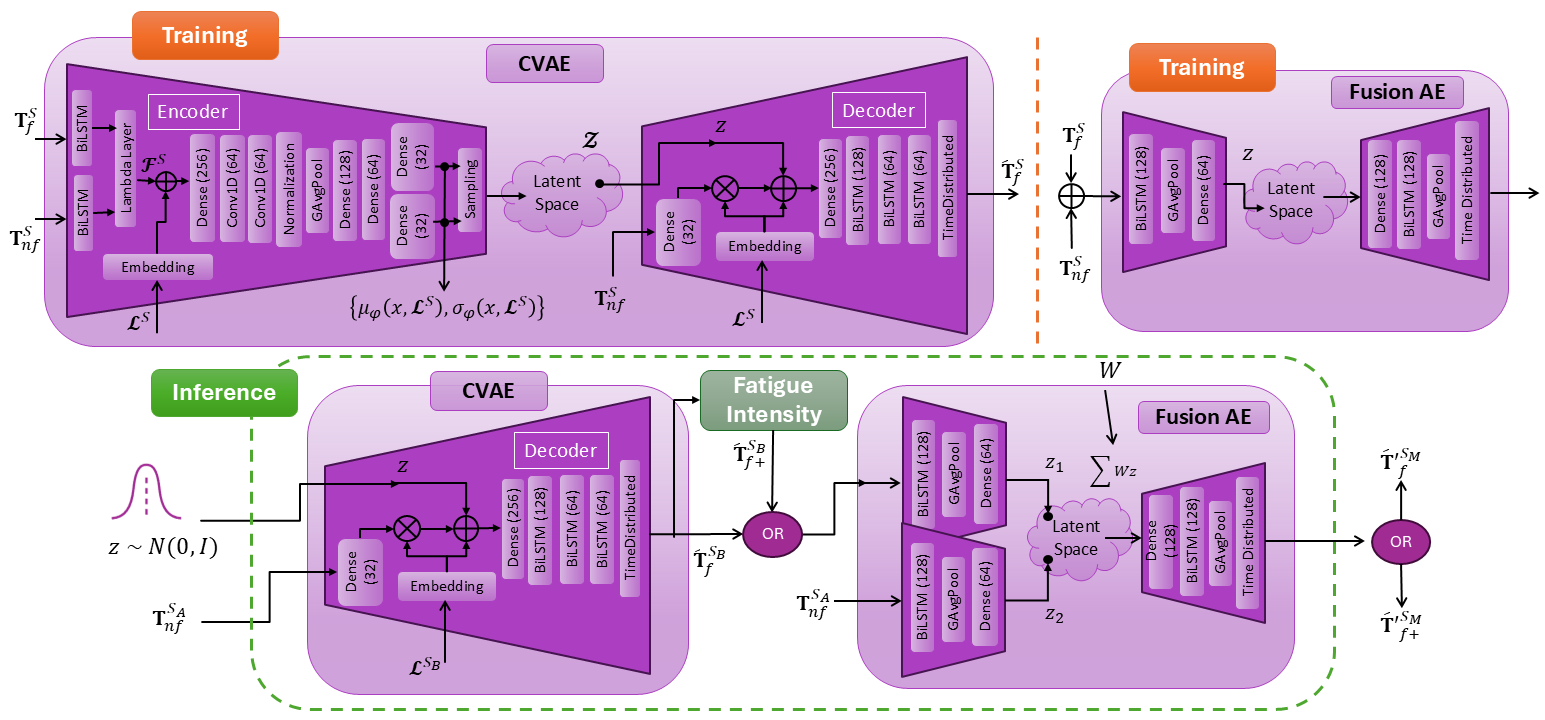}
\caption{A closer look at the architecture of CVAE and FusionAE comprising the Fatigue Features module. During training the CVAE encodes fatigue profiles in hidden unit space, each denoted as $\mathcal{F}^S$, and reconstructs fatigued joint torque sequences based on a non-fatigued torque sequence, $\mathbf{T}_{nf}^S$, a subject embedding $\mathcal{L}^S$, and a latent vector $z$ sampled from latent space, while the FusionAE is trained as a conventional autoencoder. During inference, only the decoder of CVAE is employed to generate a fatigued sequence $\hat{\textbf{T}}_{f}^{S_B}$, which can be fed to FusionAE along with a non-fatigued torque sequence of another subject (e.g. $\textbf{T}_{nf}^{S_A}$, with $S_B \neq S_A$), to produce more varied results and intermediate fatigued states. It is worth mentioning that the Fatigue Intensity module intervenes between the CVAE and AE, to accentuate the fatigue characteristics of the resulted motion. Therefore, instead of directly providing $\hat{\textbf{T}}_{f}^{S_B}$ to FusionAE, the output of the CVAE can be fed to the Fatigue intensity module to produce $\hat{\textbf{T}}_{f+}^{S_B}$, which can be given as input to the autoencoder. FusionAE will produce a fatigued torque sequence $\hat{\textbf{T}'}_{f}^{S_M}$ given the output of CVAE $\hat{\textbf{T}}_{f}^{S_B}$, or generate $\hat{\textbf{T}'}_{f+}^{S_M}$ given the output of 3CC-$\lambda$. Whether the Fatigue Intensity module is "activated" or not, can be specified by the user.}  
\label{fig:CVAE_AE}
\end{figure*}

The architecture of both the encoder and decoder of our CVAE model is depicted in Fig \ref{fig:CVAE_AE}. During training, the encoder takes as input both the non-fatigued ($\textbf{T}_{nf}$) and fatigued ($\textbf{T}_{f}$) joint torque sequences of all subjects, as well as subject embeddings ($\mathcal{L} = \{\mathcal{L}^{S}\}$), and encodes the fatigue profiles $\mathcal{F}$ of the subjects of the training dataset in latent space. Each fatigue profile is expressed as $\mathcal{F}^S = f(\textbf{T}_{nf}^S, \textbf{T}_{f}^S)$, where $f$ is basically the residual capacity $RC$, i.e., the ratio between fatigued and non-fatigued torques as arises from 3CC theory \cite{Xia2008, Cheema2023} (refer to the "Fatigue Intensity" - Section \ref{sec:fatigue_intensity} - for more details on 3CC and $RC$).

The loss of our CVAE model consists of two terms (see Eq. \ref{eq:total_loss}); the Reconstruction Loss, $L_{rec}$ which is a Mean Squared Error (MSE) loss to account for the distance between the ground-truth torques $\textbf{T}_{nf}$ and the residual torques as shown in Eq. \ref{eq:rec_loss}, and the Kullback-Leibler (KL) loss, $L_{KL}$ which describes the discrepancy between the encoder's distribution $q_{\phi}(z|x,\mathcal{L})$ and prior distribution $p_{\theta}(z|\mathcal{L})$ as described in Eq. \ref{eq:kl_loss}: 

\begin{equation}
     L = L_{rec} + L_{KL} 
     \label{eq:total_loss}
\end{equation}
\begin{equation}
    L_{rec} = \mathbb{E}_{q_{\phi}(z|x,\mathcal{L})}[MSE(\textbf{T}_{nf}^S, p_{\theta}(x|z, \mathcal{L}))]
    \label{eq:rec_loss}
\end{equation}
\begin{equation}
    L_{KL} = D_{KL} (q_{\phi}(z|x,\mathcal{L}) ||  p_{\theta}(z|\mathcal{L}))
    \label{eq:kl_loss}
\end{equation}

The reconstruction loss, $L_{rec}$, aids the decoder to learn to construct fatigued torque sequences based on vectors $z$ sampled from the latent distribution, i.e. latent fatigued features, while the second term, $L_{KL}$, compares two probability distributions, the learned distribution $q_{\phi}(z|x,\mathcal{L})$ and a simple Gaussian distribution, $p_{\theta}(z|\mathcal{L})$. The latter term acts as a regularization term that encourages the latent variables to follow a normal distribution, enabling the construction of a continuous latent space based on which novel motions can be generated. 

The model was trained in Python Keras \cite{chollet2015keras} for $50$ epochs, with a batch size of $32$ and utilizing the Adaptive Moment Estimation (Adam) \cite{Kingma2014} optimizer with an initial learning rate of 0.001. To mitigate the risk of gradient decent getting stuck on local minima, we used an adaptive learning rate scheduling technique that automatically decreases the learning rate when the learning process plateaus. In addition, we incorporated a linear warm-up method of the $L_{KL}$ term, to gradually increase its contribution to the loss function. This slow introduction prevents the KL regularization term from dominating the early stages of the training process, allowing the decoder to first develop a robust reconstruction capability before being regularized towards the latent prior. Hence, a $\beta \in [0, 0.5]$ weight is introduced in the loss function, modifying Eq. \ref{eq:total_loss} as follows:     

\begin{equation}
     L = L_{rec} + \beta L_{KL} 
     \label{eq:total_loss_2}
\end{equation}

During inference, only the CVAE decoder is employed, which is fed with a non-fatigued torque sequence $\textbf{T}_{nf}^{S_A}$, a condition, i.e. a subject embedding $\mathcal{L}^{S_B}$, where $S_A \neq S_B$, and a latent vector $z$ sampled from a Gaussian distribution, to generate a joint torque sequence, $\hat{\textbf{T}}_{f}^{S_B}= p_{\theta}(x|z, \mathcal{L})$, that incorporates fatigue features closer to the ones of subject $S_B$ (the one that the label specifies).  

\paragraph{Fusion Autoencoder}

The CVAE model leverages non-fatigued seeds, to synthesize torque sequences with fatigue characteristics to align closely with the subject‑specific fatigue profile indicated by the conditioning label. However, in contrast with the expected CVAE behavior - which should produce diverse outputs - different $z$ vectors in our case yield similar torque sequences, showing a hindrance in stochasticity. We assume that this is due to the very limited training data (torque sequences of $\sim6000$ frames stemming from $16$ subjects) that prevents the model from learning a continuous latent space from which new data samples are generated. Therefore, to compensate for this slight posterior collapse we created a simple Autoencoder (AE), named \textit{Fusion Autoencoder} (FusionAE), to enhance the variability of the Fatigue Features module. 

The FusionAE leverages latent space vector arithmetic operations influenced by facial features editing in images \cite{Radford2015}, to fuse two or more motion sequences for unique fatigue motion generation. More specifically, as depicted in Fig. \ref{fig:CVAE_AE} (bottom), this model takes either the output of the CVAE network, $\hat{\textbf{T}}_{f}^{S_B}$, i.e. an already fused fatigued torque sequence, or the output of the Fatigue Intensity module, $\hat{\textbf{T}}_{f+}^{S_B}$, and non-fatigued torques of a subject (either $S_A$ or any other subject, e.g. $S_C$ - here to avoid any confusion we will utilize the non-fatigue ground-truth of $S_A$), $\textbf{T}_{nf}^{S_A}$, and maps them in latent representations $z_1$ and $z_2$ respectively, which are fused in latent space via a weighted sum as indicated in Eq. \ref{eq:weighted_sum}, before being decoded back to torque sequences by the AE's decoder:

\begin{equation}
    \hat{\textbf{T}'}_{f}^{S_M} = \sum_{n=1}^{N}  W_nz_n
    \label{eq:weighted_sum}
\end{equation}

where $N=2$ is the number of the torque sequences being fused, $S_M \notin \mathcal{L}$, and $W$ is a vector of fusion weights for which holds $\sum W_n = 1$. These weights enable the static or dynamic fusion of fatigued and non-fatigued sequences in latent space. By "static" we refer to the manual assignment of fixed weights that deterministically fuse percentages of two torque sequences yielding motions that exhibit specific fatigued characteristics as well as producing intermediate fatigued states. In contrast, "dynamic" denotes a gradual interpolation between two fatigue profiles to create progressively fatigued motion sequences. Indicative examples are provided in the Results Section \ref{sec:results}, to further clarify the contribution of FusionAE. 

This model is trained in Keras as a classical AE, learning to reconstruct its input, and then the pretrained encoder and decoder of FusionAE are employed during inference alongside latent space vector arithmetic. MSE is used as the loss function, Adam as the optimizer, and the network is trained over $20$ epochs with a batch size of $32$.   

\subsubsection{Fatigue Intensity Module}
\label{sec:fatigue_intensity}

The Fatigue Intensity module constitutes an extension of the 3CC-$\lambda$, proposed in \cite{Loi2025}, to model the effect of fatigue accumulation in temporally evolving motion sequences. 3CC-$\lambda$ is a PINN adaptation of the Three-Compartment Controller \cite{Liu2002, Xia2008, Frey-Law2012}, a state machine that governs the transition of muscle motor units of a human limb among three compartments - active ($M_A$), fatigued ($M_F$), or resting ($M_R$) - during human motion, thereby enabling the simulation of fatigue dynamics. The sum of the percentages of muscle motor units in the three compartments forms a total of $100\%$ expressed in $\%MVC$ maximum voluntary contractions. The remaining maximum exerted joint torque due to fatigue is represented by Residual Capacity ($RC$), which as originally defined in \cite{Xia2008} is described by Eq. \ref{eq:RC}:

\begin{equation}
    RC(t) = 100\% - M_F = M_A + M_R 
    \label{eq:RC}
\end{equation}

In \cite{Loi2025}, a $\lambda \in [0, 1]$ factor was introduced in Eq. \ref{eq:RC}, to reduce the impact of fatigue in motion by a small percentage, thereby reducing abrupt motion variations and avoiding sharp frame transitions. This factor is joint-specific, to enable fatigue configurations on a joint level as well as better alignment of joint fatigue in various movements. 

\begin{equation}
    \hat{RC(t)} = 100\% - \lambda * M_F 
    \label{eq:RCC_hat}
\end{equation}

In this work, to enhance the smooth integration of fatigue in joint torques in order to enable the realistic rendering of fatigue accumulation in the residual motion, instead of expressing $RC$ as a linear decay of maximum exerted torques, we formulated $RC$ as shown in Eq. \ref{eq:RCC_hat_d}:

\begin{equation}
    \hat{RC}(t)_d = -e^{log\hat{RC(t)}} + 1
    \label{eq:RCC_hat_d}
\end{equation}

This formulation is consistent with relevant literature \cite{Pethick2015, Kent2014}, where the time-course progression of the maximum voluntary joint torque due to fatigue is reported to be non-linear, often starting with a rapid initial drop followed by a slower asymptotic decline as fatigue accumulates.

Subsequently, unlike our previous work \cite{Loi2025} where $RC$ is utilized to inject fatigue in non-fatigued sequences, $RC$ is applied to already fatigued joint torques supplied by the Fatigue Features module (i.e. CVAE), as a time-varying factor to model the progressive decline in maximum joint torque capacity, yielding $\hat{\textbf{T}}_{f+}^{S_B}$ (Eq. \ref{eq:3CC-lambda}):   

\begin{equation}
    \hat{\textbf{T}}_{f+}^{S_B} = \hat{RC}_d * \hat{\textbf{T}}_{f}^{S_B} 
    \label{eq:3CC-lambda}
\end{equation}

The $u$ factor given as input to the Fatigue Intensity module as shown in Fig. \ref{fig:fatigue_fusion}, acts as a lower limit for $RC$ (or upper limit for $\lambda * M_F$, respectively), resulting in $RC$ guiding the progressive accumulation of fatigue up to a specific level (e.g. $60\%$).  

The architecture, loss function and training procedure of 3CC-$\lambda$ described in \cite{Loi2025}, were also followed in this work. 

\subsubsection{Post-Processing}

An algorithm similar to the one used during training data post-processing (see "Post-Processing" Subsection in Appendix), was implemented to better align root motion in fatigued generated movements and mitigate foot skating artifacts. A Savitzky-Golay filter was also utilized to smooth out any residue motion artifacts.

\section{Results}
\label{sec:results}

\subsection{Dataset \& Data Pre-processing}
\label{sec:dataset}
The proposed framework was trained and evaluated upon non-fatigued and fatigued gait motion sequences stemming from the DUO-Gait \cite{DUO_GAIT} dataset. The dataset contains IMU sensor data (tri-axial acceleration and angular velocity) from 16 healthy adults (aged $21$ to $25$, $50-50$ analogy of female-male), who performed $6$ minutes of non-fatigued walking, $6$ minutes of fatigue protocol (sit-to-stand), and $6$ minutes of fatigued walking. The gyroscope and accelerometer data were fused to derive orientation data, i.e. data describing each IMU sensor's rotational movement in 3D space, described in Quaternions ($\{w, x, y, z\}$), provided also in the dataset. The experimental setup that the authors utilized to infer sensor measurements, includes $9$ synchronized IMU sensors, namely, one sensor on the head, chest, and lower back, and a pair of sensors for the wrists, legs (placed under the knee - tibia bone), and feet, with a sampling rate of 128 Hz. These sensor readings are available for both single (only gait) and dual task (i.e. walking along with realising a cognitive task such as a mathematical operation) gait, resulting in the recording of $4$ walking sessions per subject, namely under single-task non-fatigue, single-task fatigue, and their respective dual-task conditions. The IMU data were recorded in a single $18$-minute long file for each subject and per state (e.g., single-task non-fatigued, dual-task fatigue, etc.), which was also provided segmented into three $6$-minute parts containing each one of the aforementioned states. The latter data files are in CSV format, which can be easily manipulated through Python scripting.  

Apart from IMU measurements, spatio-temporal gait parameters such as stride length and clearance, stride and stance times, etc., extracted from gait events, are also provided in the DUO-Gait dataset. Aggregated gait parameters in terms of mean, coefficient of variation, and symmetry index were computed per subject and walking condition. Moreover, the dataset includes demographic and physiological variables 
along with transcripts of the subject responses during the cognitive task, to facilitate subsequent research and analyses. Along with the DUO-Gait dataset, a Python code repository was made publicly available, containing functions to enable the segmentation of IMU data and computation of the residual gait parameters, as well as their corresponding aggregated parameters. In this work, we developed our own Python scripts to process and make the best use of this dataset, meeting the needs of this framework.  

Therefore, since most motion synthesis and physics-based modeling techniques rely their results on joint position or angle trajectories, we designed a data pre-processing pipeline to extract motion sequences from orientation data for both non-fatigued and fatigued motions. More specifically, the transformation between experimental orientation data (in the form of Quaternions) into joint angle sequences (Euler Angles) was facilitated by OpenSense \cite{Al2022}, an auxiliary open-source tool integrated into OpenSim (Version 4.5). In particular, OpenSense estimates joint kinematics of the lower limbs, based on IMU signals, while addressing error accumulation issues, which hinder the accuracy of long-term motion estimation.


\comm{As illustrated in \red{Figure XX},} 
\comm{The steps towards extracting clean and smooth non-fatigued/fatigued movements from IMU measurements, are described in the subsections below.} 
The data-preprocessing steps revolve mainly around Data Segmentation and Calibration techniques, while data interpolation and post-processing methods were utilized to produce accurate motion sequence (i.e. joint angle) estimations. 
More details regarding the data modulation methods applied on the experimental IMU orientation data, are provided in the Appendix, which accompanies this work as supplemental material.

\subsection{Fatigued Animation}
\label{sec:fatigued_animation}

Our framework's capability to effectively produce full-body fatigued motions incorporating various fatigued characterists, was evaluated through three sets of experiments and respective comparisons with experimental ground-truth data; \textit{fatigue profile transfer}, \textit{fatigue profile fusion} and \textit{generation of progressively fatigued motions}. Throughout this section, subject annotations (e.g. A, B, C.., etc.) are example/figure specific and do not necessary refer to the same subject (fatigue profile) each time. 

The animation results were recorded in Unity Game Engine environment \cite{Unity}, utilizing the 3D humanoid avatar from the \textit{Starter Assets: Character Controllers | URP}\footnote{Available online at: \url{https://assetstore.unity.com/packages/essentials/starter-assets-character-controllers-urp-267961}} package developed by Unity Technologies and open-source distributed in the Unity Asset Store. 

\subsubsection{Fatigue Profile Transfer}

\comm{"
Experiment 1: 
Experiment 2: The framework takes again as input the non-fatigued motion of subject A and produces a residual fatigued motion closer to the one of subject B. The fatigue tempo of subject A is retained, while no additional fatigue is added from the Fatigue Intensity module.
Experiment 3:}

\begin{figure}[h!]
\centering
\includegraphics[width=\columnwidth]{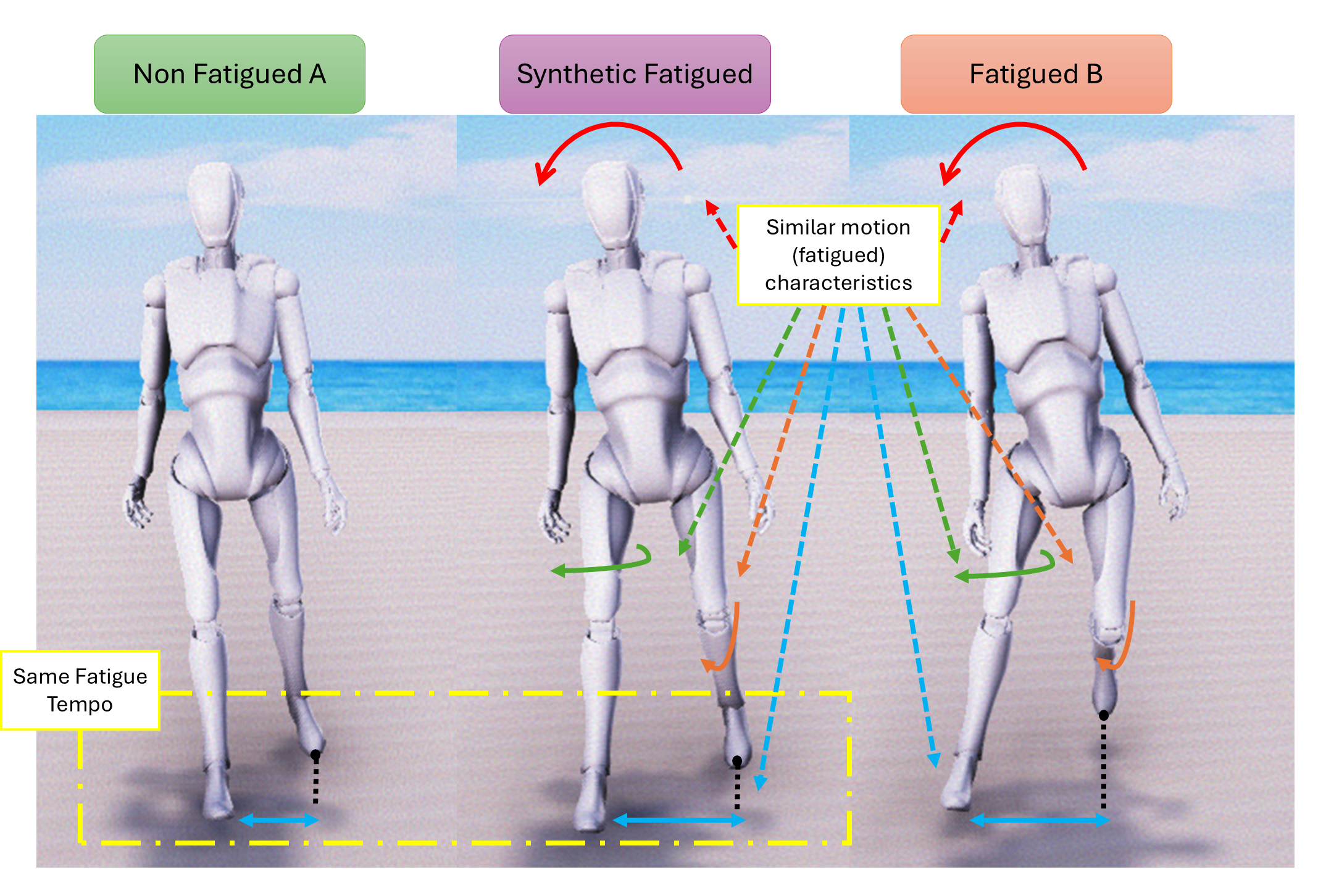}
\caption{Non-fatigued A: The ground-truth non-fatigued gait motion of subject A; Synthetic Fatigue: The output of FatigueFusion framework; Fatigued B: The ground-truth fatigued gait motion of subject B. The red arrows denote the lumbar bending angle, the green arrows the hip rotation angle, the orange arrows the knee flexion angle, and the light blue arrows indicate the step width. Overall, the residual fatigued motion resembles the fatigue features of B, while has the tempo of A. All poses are in the same frame.}  
\label{fig:features_B_tempo_A}
\end{figure}

\begin{figure}[h!]
\centering
\includegraphics[width=\columnwidth]{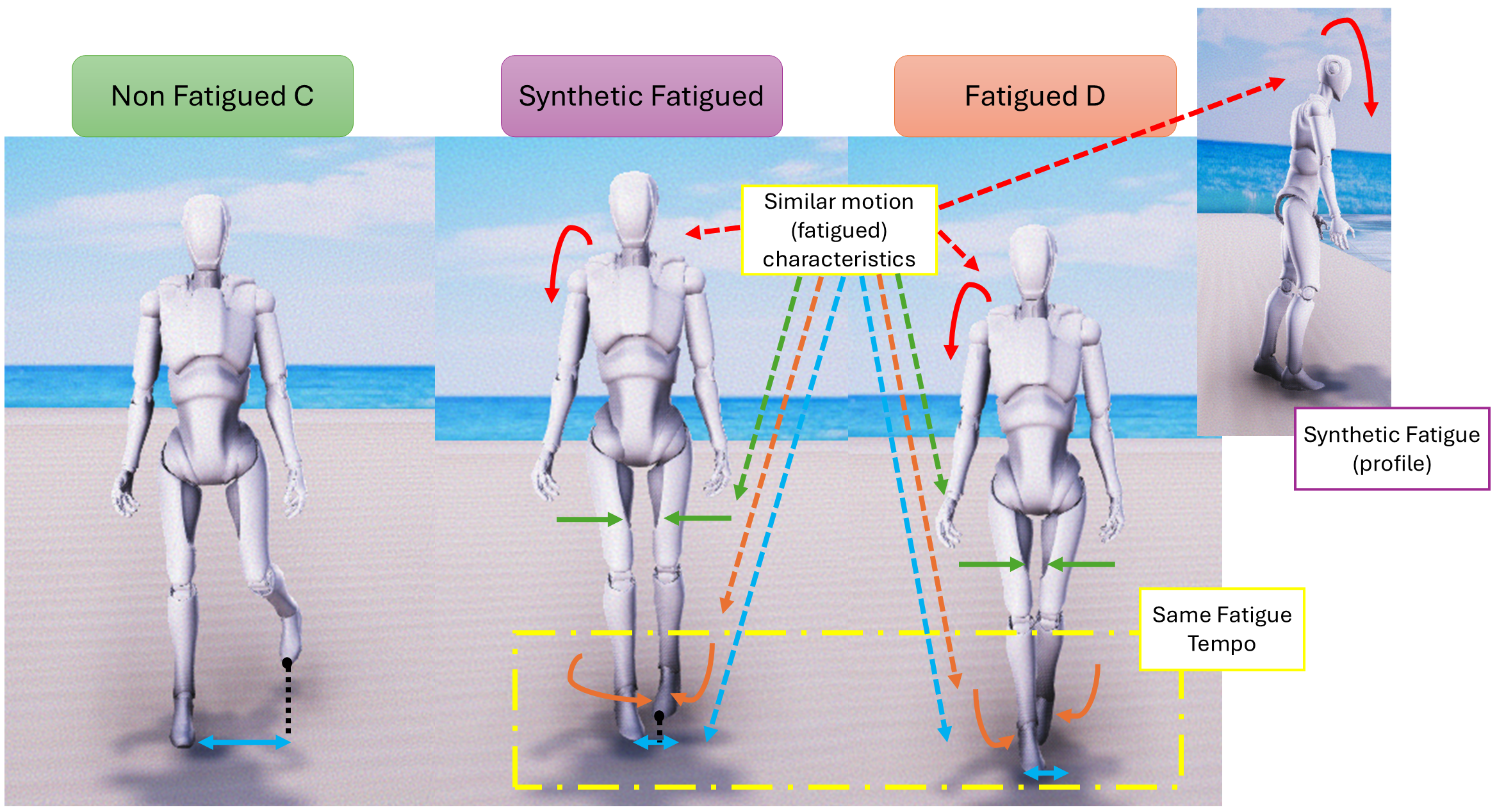}
\caption{Non-fatigued C: The ground-truth non-fatigued gait motion of subject C; Synthetic Fatigue: The output of FatigueFusion framework; Fatigued D: The ground-truth fatigued gait motion of subject D. The red arrows annotate the lumbar extension angle, the green arrows the hip adduction angle, the orange arrows the subtalar angle, and the light blue arrows indicate the step width. The depicted synthesized motion exhibits both spatial and temporal fatigue features similar to the ones of the fatigued motion of subject D. All poses are in the same frame.}  
\label{fig:features_D_tempo_D}
\end{figure}

The capability of the framework to transfer subject-specific spatial and temporal fatigued features, was assessed in this experiment. Indicatively, in Figures \ref{fig:features_B_tempo_A} and \ref{fig:features_D_tempo_D}, we present two cases of fusing personalized fatigue characteristics into non-fatigued movements. In Fig. \ref{fig:features_B_tempo_A} the framework is fed with the non-fatigued motion of subject A, $\mathbf{Q}_{nf}^{S_A}$, and outputs a fatigued motion with fatigued spatial characteristics resembling the ones of subject B (i.e. fatigue profile of B), yielding $\mathbf{\hat{Q'}}_{f+}^{S_M}$. Similar fatigue features to the ground-truth fatigue motion of subject B that the synthesized motion has, constitute, decreased hip rotation and adduction, which cause the step width to increase, increased lumbar bending and knee flexion angle. The temporal characteristics of subject A are maintained. A $20\%$ more fatigue is injected from the Fatigue Intensity module. Likewise, in Fig. \ref{fig:features_D_tempo_D}, the fused fatigue motion carries fatigue features and tempo similar to those of subject D's fatigued gait, while $30\%$ of fatigue was added to the residual motion. Subject D presents decreased lumbar extension and also stumbles while walking fatigued, which is transferred to motion as increased hip adduction and subtalar angle resulting in decreased step width and length, features also adopted in the synthetic motion. 

\subsubsection{Fatigue Profile Fusion}

\comm{
-> lumbar extension -> 23 
-> hip adduction r -> 08
-> hip flexion r -> 07
-> knee angle r -> 10
-> ankle angle r -> 12

-> a fatigue profile for each subject} 

Fatigue profile fusion experiments were conducted to assess the ability of the proposed model to generate diversified fatigued motion sequences, while providing no fatigued data as input. This process involves fusing two or more fatigue profiles together, i.e. personalized fatigued features such as jazzy leg, extreme lumbar bending, etc., to produce novel unique fatigued motions. To do so, we provided a non-fatigued motion and various combinations of control parameters as input to our framework, which yielded the corresponding results. 

\begin{figure}[h!]
\centering
\includegraphics[width=\columnwidth]{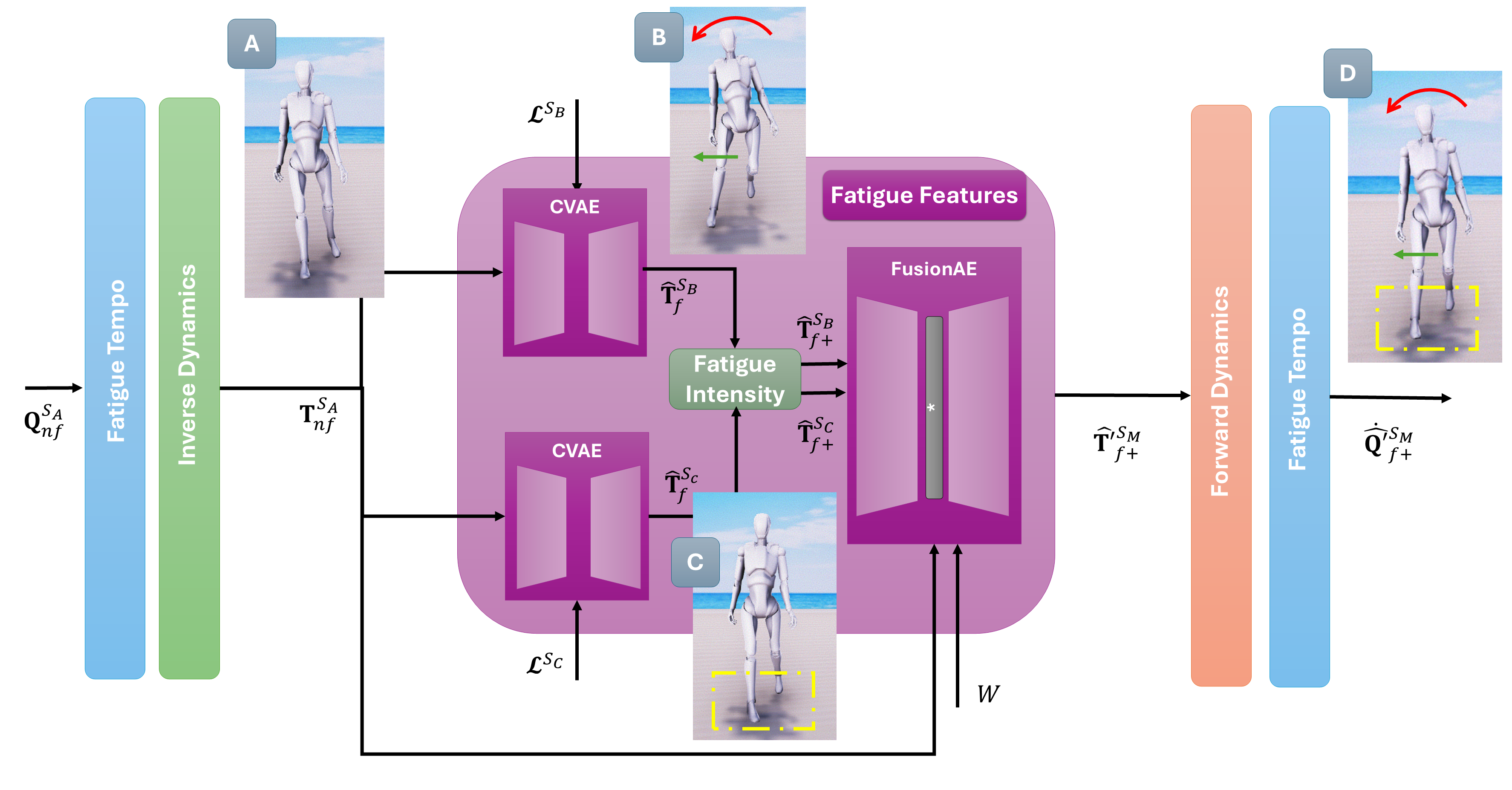}
\caption{Figure A: Ground-truth non-fatigued gait of subject A; Figure B: Output of CVAE given non-fatigued torques of subject A, $\mathbf{T}_{nf}^{S_A}$, and the label corresponding to subject B, $\mathcal{L}^{S_B}$. This motion is closer to the fatigue profile of B; Figure C: Output of CVAE given the same input, $\mathbf{T}_{nf}^{S_A}$, and different subject embedding, the one of subject C, $\mathcal{L}^{S_C}$. The resulting motion carries fatigue characteristics similar to the ones of subject C's fatigue profile; Figure D: The framework's output, incorporating fatigue features from both fatigue profiles. The yellow box indicates subtalar angle, while the red and green arrows denote the pelvis tilt and hip adduction angle.}  
\label{fig:fatigue_profile_fusion}
\end{figure}

In Fig. \ref{fig:fatigue_profile_fusion}, we present our findings while applying the fatigue characteristics of the fatigued gait of subject B and C on the non-fatigue movement of subject A. As illustrated in this figure, we may deploy the CVAE component twice, given the same input, non-fatigued torques $\mathbf{T}_{nf}^{S_A}$, and different subject embeddings, namely $\mathcal{L}^{S_B}$ and $\mathcal{L}^{S_C}$, to produce fatigue torques incorporating features from subjects B and C, respectively. These generated torque sequences are then fed to the FusionAE to produce novel fatigue motion that exhibits fatigue characteristics similar to those of \textit{both} the fatigue gait of subjects B and C. As illustrated in \ref{fig:fatigue_profile_fusion}, the most distinct fatigued features that the residual fatigued motion, $\mathbf{\dot{\hat{Q'}}}_{f+}^{S_M})$, presents, are decreased hip adduction and increased pelvis tilt similar to the fatigue profile of subject B, and increased subtalar angle resembling the fatigue profile of subject C. 

\comm{During this experiment, the Fatigue Features module was mainly employed as illustrated in Fig. \ref{fig:fatigue_fusion}. More specifically, non-fatigued torques, ..., were fed to the ....} 
By entangling more fatigued features, i.e mapped temporal characteristics per stance phase (e.g. different pacing due to fatigue) from the Fatigue Tempo module, and various fatigue levels using the Fatigue Intensity component, more variance is rendered on the resulting motion.

\subsubsection{Progressively Fatigued Motions}

\begin{figure*}[h!]
\centering
\includegraphics[width=\textwidth]{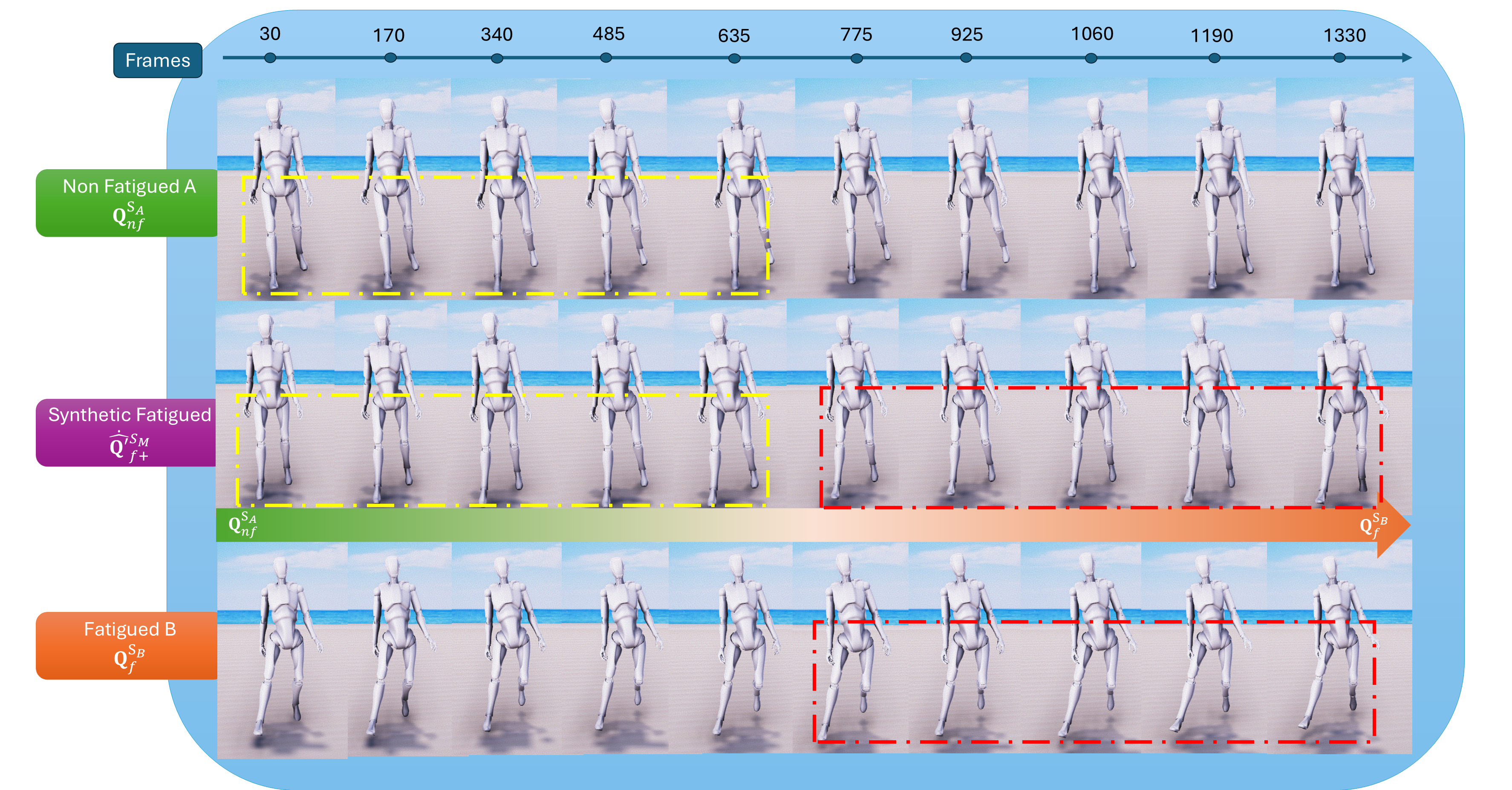}
\caption{The transition from the non-fatigued motion of subject A, $\mathbf{Q}_{nf}^{S_A}$, to a fatigued motion exhibiting the personalized fatigued gait features of subject B, $\mathbf{\hat{Q'}}_{f+}^{S_M}$. The ground-truth fatigued motion of subject B, $\mathbf{Q}_{f}^{S_B}$, is also provided.}  
\label{fig:progressive_fatigue_animation_results}
\end{figure*}

\begin{figure*}[h!]
\centering
\includegraphics[width=\textwidth]{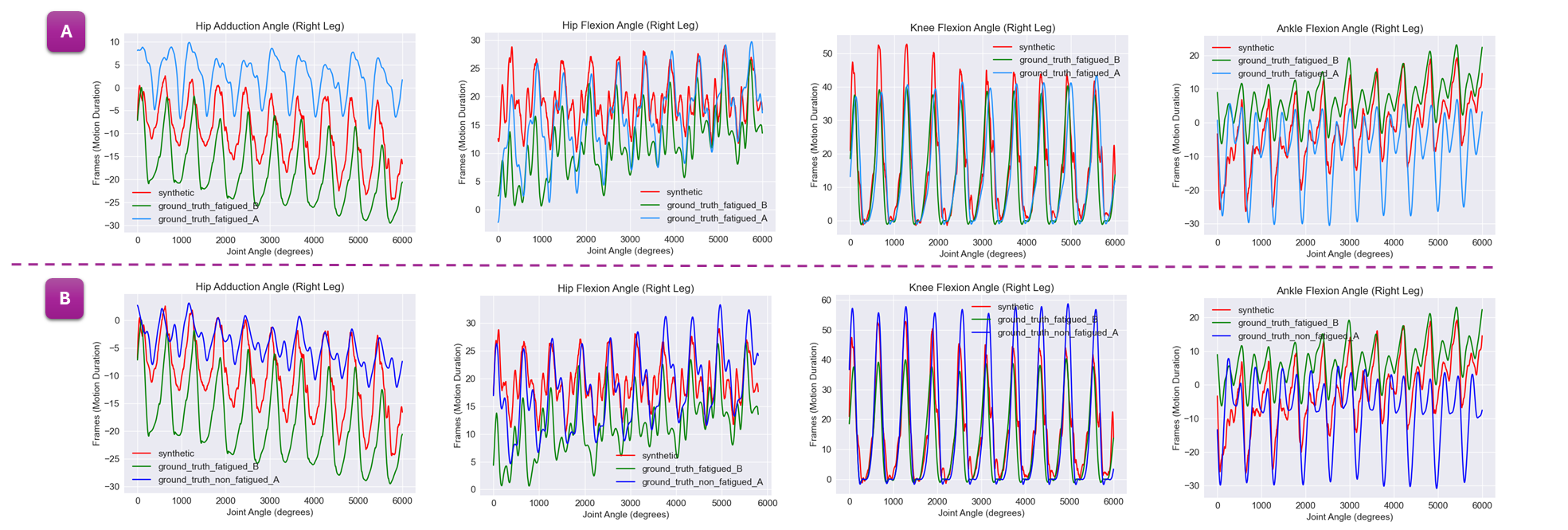}
\caption{Comparison of hip adduction, hip, knee and ankle flexion angles, across A) synthesized (output of FatigueFusion model - red line), ground-truth fatigued gait motion of subject A (light blue line) and that of subject B (green line), and B) synthesized (red line), ground-truth non-fatigued motion of subject A (blue line) and ground-truth fatigued motion of subject B (green line).}  
\label{fig:progressive_fatigue_results}
\end{figure*}

Figure \ref{fig:progressive_fatigue_animation_results} showcases the capability of the model to generate progressively fatigued motion sequences. In this experiment, the non-fatigued motion of subject A is gradually getting fatigued while acquiring the fatigued spatial and temporal features of subject B. An additional $30\%$ of fatigue is injected to make the fatigued residual motion more evident. Indicatively, in Fig. \ref{fig:progressive_fatigue_animation_results} we annotate the lower body of the avatar, to show that by the beginning of the motion, the synthetic fatigued sequence retains to a certain extend the gait features of the non-fatigued ground-truth motion of subject A, while after some frames, it obtains the fatigue characteristics of the ground-truth fatigue motion of subject B. For example, subject B seems to be dragging their right leg while walking in a fatigued state, which is the result of increased lumbar bending and  decreased hip adduction and knee flexion angle as is also apparent in Fig. \ref{fig:progressive_fatigue_results}. In this figure depicting lower body joint angles, we observe that, over the course of time, the synthesized motion (red line) reproduces motion patterns and converges to values of the ground-truth fatigued motion of subject B (green line) as also deduced from the animation results (Fig.\ref{fig:progressive_fatigue_animation_results}).    

A similar gradual transition from non-fatigued to fatigued state, i.e. fatigue accumulation, can be achieved via exploiting the Fatigue Intensity module's capabilities, alongside the dynamic fusion of fatigue profiles in latent space. Furthermore, FatigueFusion can be further employed to produce intermediate fatigue states, by statically fusing fatigue features into non-fatigued sequences, in latent space. For instance, intermediate frames from Fig. \ref{fig:progressive_fatigue_animation_results}, can be obtained by simply modulating fusion weights, $W$.

\comm{++ As it is evident from figures ... Intermediate Fatigue States ++ 
-> experiment with different fusion weights and create an image for that} 

\subsection{Quantitative Analysis and Comparisons}
\label{sec:quantitative_analysis}



\begin{table}[ht]
\setlength\tabcolsep{1.5pt}
\centering
\begin{threeparttable}
\caption{Comparison between Generated and Ground-Truth Motions}
{\begin{tabular}[t]{ccccccccc}
\toprule
Ex No. & \multicolumn{3}{c}{Fatigue Features} & \multicolumn{4}{c}{Evaluation Metrics} \\  
\midrule
& FSF & FTF & FI & $MAE_{GT_{A}}\downarrow$ & $R^2_{GT_{A}}\uparrow$ & $MAE_{GT_{B}}\downarrow$ & $R^2_{GT_{B}}\downarrow$ \\
\midrule
I & $\rightarrow$A & $\rightarrow$A & $20\%$ & $2.50^{\pm0.008}$ & $0.81^{\pm0.009}$ & $4.72^{\pm0.005}$ & $0.67^{\pm0.002}$\\
II & $\rightarrow$A & $\rightarrow$A & $0\%$ & $2.41^{\pm0.003}$ & $0.85^{\pm0.002}$ & $4.68^{\pm0.006}$ & $0.70^{\pm0.009}$\\
III & $\rightarrow$B & $\rightarrow$A & $0\%$ & $3.61^{\pm0.004}$ & $0.71^{\pm0.005}$ & $2.23^{\pm0.008}$ & $0.85^{\pm0.009}$\\
IV & $\rightarrow$B & $\rightarrow$B & $30\%$ & $3.80^{\pm0.004}$ & $0.69^{\pm0.005}$ & $2.16^{\pm0.005}$ & $0.88^{\pm0.002}$\\
V & $\rightarrow$B & $\rightarrow$B & $0\%$ & $3.69^{\pm0.002}$ & $0.71^{\pm0.001}$ & $2.18^{\pm0.003}$ & $0.90^{\pm0.009}$\\
VI & $\rightarrow$A & $\rightarrow$B & $20\%$ & $2.16^{\pm0.008}$ & $0.77^{\pm0.008}$ & $4.80^{\pm0.004}$ & $0.66^{\pm0.002}$\\
\bottomrule
\end{tabular}}
\label{tab:comparison}
\begin{tablenotes}
\item[$^a$] {Comparison between fatigued fused motion sequences produced from our framework and corresponding fatigued ground-truth motions from DUO-Gait Dataset. FSF: Fatigue Spatial Features; FTF: Fatigue Temporal Features; FI: Fatigue Intensity; Ex. No: Experiment Number; $GT_A$: Ground truth fatigued motion of subject A (with label $S_A$); $GT_B$: Ground truth fatigued motion of subject B (with label $S_B$). The $\rightarrow$ is a notation for "closer" (e.g. FSF $\rightarrow$B annotates a motion whose fatigue spatial features are closer to the ones of the fatigued motion of subject B), while $\uparrow$ and $\downarrow$ connote that higher and lower values, respectively, indicate better results.}
\end{tablenotes}
\end{threeparttable}
\end{table}%

To further demonstrate the validity of our results presented in the previous Section \ref{sec:fatigued_animation}, we employ Mean Absolute Error (MAE) and Pearson's Correlation Coefficient ($R^2$) to measure how close and correlated are the generated fatigued sequences with the corresponding ground-truth motions from DUO-Gait dataset. Table \ref{tab:comparison}, summarizes our results over $4$ experiments, where fatigued motions incorporating different spatial and temporal fatigue characteristics are produced. In particular, we report the following experiments: 
\begin{enumerate}
	\item{Experiment I:} A motion generated from our FatigueFusion framework exhibiting spatial and temporal fatigue features resembling the corresponding fatigue features of subject A (ground-truth fatigue motion), while incorporating additional $20\%$ fatigue from the Fatigue Intensity module.
	\item{Experiment II:} The same as Experiment I, without considering additional fatigue.
	\item{Experiment III:} A synthetic motion incorporating the fatigue profile of subject B, fatigue tempo of A, and no additional fatigue.
	\item{Experiment IV:} A synthetic motion having spatial and temporal fatigue features more close to the corresponding fatigue features of subject B, while $30\%$ fatigue was added.
	\item{Experiment V:} The same as Experiment IV, without considering additional fatigue.
	\item{Experiment VI:} A synthetic motion incorporating the fatigue profile of subject A, fatigue tempo of B, and $20\%$ additional fatigue.
\end{enumerate}

Smaller MAE and values closer to $1$ for $R^2$ indicate better results. 

From Table \ref{tab:comparison}, we observe that motion sequences exhibiting fatigue features closer to the ones of a subject (e.g. $S^B$ for experiments III and IV), tend to be more correlated (higher $R^2$) and similar (lower MAE) with the respective ground-truth fatigue motion (e.g. $\mathbf{Q}_{f}^{S_B}$ - marked as $GT_B$ in this table). Additional fatigue injected from the Fatigue Intensity module, slightly increases MAE and reduces $R^2$, as it adds variance to the residual motions, differentiating them from corresponding ground-truth motions, as also shown in the Ablation Study conducted in Section \ref{sec:ablation}. 

\begin{figure*}[h!]
\centering
\includegraphics[width=\textwidth]{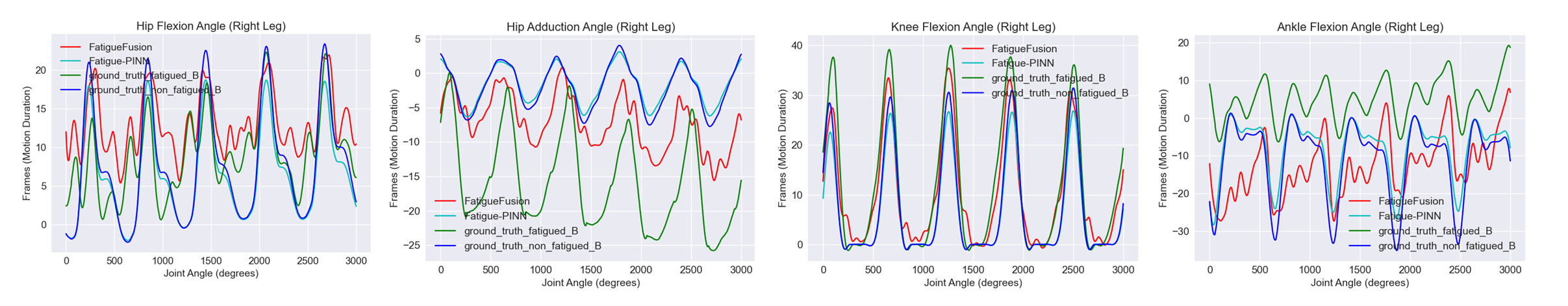}
\caption{Comparison of Fatigue-PINN results with the ones of the proposed FatigueFusion. Both frameworks are fed with the non-fatigued motion of subject B as input, while $40\%$ of fatigue was added. Fatigue Fusion produces motions with characteristics similar to the fatigued ground-truth.}  
\label{fig:fatigue_pinn_vs_fatigue_fusion}
\end{figure*}

   
We also present a brief comparison with our previous work, Fatigue-PINN \cite{Loi2025}, where a deterministic approach to model fatigue accumulation in motions, without prior fatigue knowledge, was implemented. Figure \ref{fig:fatigue_pinn_vs_fatigue_fusion}, illustrates non-fatigued and fatigued hip flexion, hip adduction, knee and ankle flexion joint angles. It is clear, that Fatigue-PINN (light blue line) produces fatigued joint angles closer to the ones of the non-fatigued ground-truth motion, since it does not take fatigue characteristics into consideration. On the contrary, FatigueFusion (red line) can produce fatigue motion sequences resembling the motion patterns of the experimental fatigued motions, since it encodes fatigue profiles in latent space. 

\subsection{Ablation Study}
\label{sec:ablation}

We ablate the individual components of FatigueFusion to validate our proposed architecture and further assess the framework's capability of producing diverse fatigued motions. To do so, we employ the \textit{Frechet Inception Distance} (\textit{FID}) and \textit{Diversity} 
metrics, which are widely used to evaluate generative (e.g. GANs) and probabilistic data-driven models \cite{Li2025, Kang2025, Tashakori2025}. FID measures the distance between the feature distribution of ground-truth motions and the one of the residual fatigued motions, hence applied on feature vectors from generated and real motion sequences. For the Diversity metric, we utilize the formula defined in \cite{Guo2020}, to compute the average pairwise Euclidean distance between latent motion feature vectors randomly sampled from a set of all fatigued fused motions (our framework's output), to measure the variability of generated sequences across all motions (subjects). 
Smaller FID values indicate better overall quality of results, whereas higher Diversity 
values suggest more diversified motion outputs.

\begin{figure}[h!]
\centering
\includegraphics[width=\columnwidth]{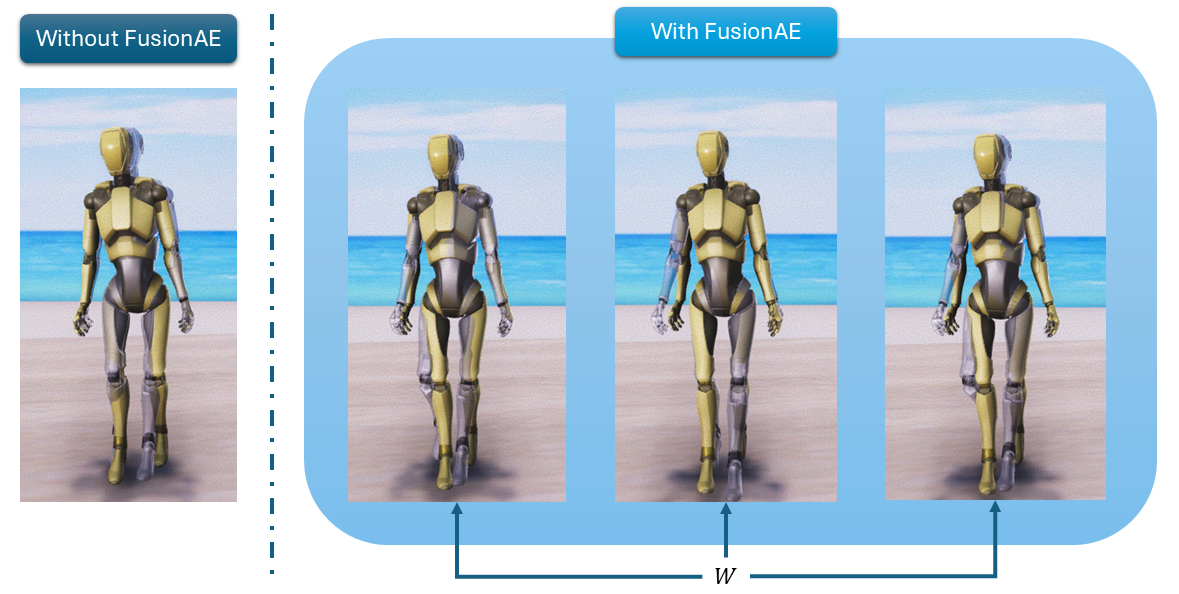}
\caption{Qualitative comparison of the output of the FatigueFusion framework with and without incorporating FusionAE in its architecture. The yellow model performs a synthetic movement with the fatigue profile of subject C,  generated by our framework, while the gray (ghost) model has the ground-truth fatigue motion of C. FusionAE aids in generating variations of CVAE's output, improving the framework's capability to generate diversified fatigued movements.} 
\label{fig:fusion_ae_results}
\end{figure}

As derived from Table \ref{tab:ablation}, the addition of FusionAE significantly lowers FID and increases Diversity metrics, indicating improved realism and diversity on generated fatigued motions, rendering this component essential for the framework. The latter further highlights the inability of CVAE to learn a fully continuous latent space, due to limited training data, thus slightly collapsing to similar motions, which leads to shrinking the covariance of generated features and ultimately increasing FID. Qualitative results of employing FatigueFusion with and without incorporating FusionAE, are illustrated in Fig. \ref{fig:fusion_ae_results}, where it is evident that the AE further perturbs animation results, hence enhancing the variability of the overall architecture. 

Adding 3CC-$\lambda$ (Fatigue Intensity) and Fatigue Tempo modules, further aids the framework to produce a wide range of motions, which is apparent even in cases where only CVAE is employed alongside them. Moreover, our framework achieving the best FID (Model Version VI), connotes that synthesized fatigued motion embeddings cluster near the real ones, meaning that generated motions look physically plausible and human-like.

\begin{table}[ht]
\setlength\tabcolsep{2.5pt}
\centering
\begin{threeparttable}
\caption{Ablation Study on FatigueFusion framework}
{\begin{tabular}[t]{ccccccc}
\toprule
Model Version & \multicolumn{4}{c}{Components} & \multicolumn{2}{c}{Metrics}\\
\midrule
 & CVAE & AE & 3CC-$\lambda$ & Tempo & FID$\downarrow$ & Diversity$\uparrow$\\
\midrule
Ground-Truth Data & - & - & - & - & $\sim0.001$ & $9.670$\\
\midrule
I & \checkmark & - & - & - & $0.214$ & $8.264$  \\
II & \checkmark & \checkmark & - & - & $0.086$ & $8.840$ \\
III & \checkmark & \checkmark & - & \checkmark & $0.072$ & $8.952$\\
IV & \checkmark & - & \checkmark & - & $0.130$ & $8.440$ \\
V & \checkmark & \checkmark & \checkmark & - & $0.059$ & $9.121$  \\
VI (FatigueFusion) & \checkmark & \checkmark & \checkmark & \checkmark & $\textbf{0.045}$ & $\textbf{9.375}$  \\

\bottomrule
\end{tabular}}
\label{tab:ablation}
\begin{tablenotes}
\item[$^b$] {The best values are marked in bold. The $\downarrow$ and $\uparrow$ arrows indicate that, respectively, lower and higher values are better.}
\end{tablenotes}
\end{threeparttable}
\end{table}%

\section{Discussion}



In this work, we present FatigueFusion, a framework that, unlike recent fatigue-driven animation works, considers spatial and temporal fatigue features, alongside the level of fatigue to produce diversified fatigued motion sequences. Especially, it incorporates three modules, namely: i) the Fatigue Features module combining a CVAE and a conventional AE enabling the extraction and fusion of individualized fatigue characteristics in latent space, ii) the Fatigue Tempo module that leverages sophisticated interpolation methods to embded/extract per stance temporal fatigue features into/from a common normalized temporal space, and iii) the Fatigue Intensity module that exploits physics-based deep neural networks to amplify the effect of fatigue in residual motions. Overall, the framework accounts for the latent entanglement of distinct fatigue gait characteristics (fatigue profiles) from various subjects, e.g. fusion of stance durations of subject A with leg dragging from the fatigue state of subject B, and stumbling of subject C. 
This encoding of fatigue profiles, inherently renders our model \textit{fatigue-specific}, differentiating our framework from current motion synthesis works, which neglect the impact of fatigue on motion, hence modelling a never-ending active state of a humanoid character. 




Three sets of experiments and comparisons with experimental fatigued data stemming from the DUO-Gait dataset, were conducted to showcase the capabilities of the model. Our findings indicate that our model is able to transfer personalized fatigue features to non-fatigued gait motions, fuse two or more fatigue profiles to produce diversified fatigued motions, and statically or gradually transition from non-fatigued to fatigued states. As mentioned above, all motion modulation tasks are performed between latent representations of fatigue profiles, facilitating the integration of this framework with any animation or simulation system, relying solely on non-fatigued motions and user-defined control parameters as input. Our quantitative analysis further confirms the visual animation outcomes, while our ablation study validates the proposed architecture and its ability to produce varied fatigued motions. 




The very limited training data poses a limitation for our framework, hindering our probabilistic model's (CVAE) ability to learn a fully continuous latent space and further affecting the generalization capability of the overall pipeline to more action classes (apart from gait) and fatigue profiles. Open-source datasets incorporating fatigue human motion sequences are scarce in literature, with most recent works providing healthy single \cite{Zafra2025} or dual-task \cite{Liao2025} gait data, and those available are either distributed upon request or contain a small amount of fatigued motion data. As a resolution, we enhanced our fatigue features fusion approach with an Autoencoder that leverages latent space vector arithmetic operations to fuse two or more motion sequences in latent space for unique fatigue motion generation, thereby further enhancing the variability of the resulting motions. Furthermore, the scarcity of datasets containing fatigued human motions in combination with the absence of frameworks considering personalized fatigue features in motion synthesis and prediction tasks, render us unable to do further comparative tests.



FatigueFusion paves the way for fatigued-driven animation incorporating various personalized fatigued features. Therefore, it does not only allow for the development of realistic animation frameworks rendering both the fatigue and active human state, but also that of biomechanics-based modeling tools. The latter can be utilized for subject-specific analysis of the impact of fatigue on an individual's posture and motion as well as design of personalized fatigue mitigation strategies and injury prevention methods. As future work, we intend to extend our model to be fully biomechanics-aware, by incorporating physics-based domain knowledge in all its components, to account for dynamic simulations instead of solely modelling the effects of externally perceived fatigue in motion patterns.

\section{Conclusion}

FatigueFusion extracts temporal and spatial fatigue features from subject-specific distributions and fuses them alongside fatigue scalings to enable the generation of a variety of unique fatigue motions, intermediate fatigued states, and progressively fatigued movements. This incorporation of fatigue-specific features in residual motions, marks a novel contribution over state-of-the-art approaches that focus on imitating the impact of fatigue accumulation in movements, without accounting for diversified fatigued-specific characteristics. The three modules constituting the FatigueFusion framework, namely Fatigue Features, Fatigue Tempo and Fatigue Intensity, contribute in fusing latent representations of fatigue features to effectively produce unique fatigue movements, without relying on fatigued motion data during inference. The framework was tested across different motion modulation tasks, such as fatigue profile transfer and fusion, with our results indicating the validity of our method.

\bibliographystyle{IEEEtran}
\bibliography{bibliography}

\newpage

\vfill

\end{document}